\documentclass[12pt,manuscript,psfig,epsfig,pdffig,superscriptaddress,floatfix,nofootinbib]{revtex4}
\usepackage[latin9]{inputenc}
\usepackage{amsmath}
\usepackage{amssymb}
\usepackage{graphicx}
\usepackage{epstopdf}
\usepackage{pstricks}
\usepackage{color}

\makeatletter
\@ifundefined{textcolor}{}
{%
	\definecolor{BLACK}{gray}{0}
	\definecolor{WHITE}{gray}{1}
	\definecolor{RED}{rgb}{1,0,0}
	\definecolor{GREEN}{rgb}{0,1,0}
	\definecolor{BLUE}{rgb}{0,0,1}
	\definecolor{CYAN}{cmyk}{1,0,0,0}
	\definecolor{MAGENTA}{cmyk}{0,1,0,0}
	\definecolor{YELLOW}{cmyk}{0,0,1,0}
}

\usepackage{dcolumn}
\usepackage{bm}
\usepackage{amsfonts}
\usepackage{cancel}

\makeatother

\begin{document}
	
\title{Axial vector current anomaly problem without regularization in Dyson scheme}


\author{Shou-Shan Bao, Shi-Yuan Li, Zong-Guo Si\\
Institute of Theoretical  Physics, Shandong University, Jinan 250100, P. R. China}







\begin{abstract}
      The loop  momenta of a single Feynman diagram in momentum space  can  be  assigned  unambiguously within the 'Dyson scheme' without referring to  the other Feynman diagrams in the complete set  to some order of coupling constant for the certain process. This fact and  scheme which were provided in Dyson's  original   paper are applied to a typical relevant problem, i.e.,   the  triangle diagrams of the  'axial vector current anomaly'.  The calculation is done in four-dimension Minkowski space-time straightforwardly  without the aid of any  regularization.  The   linearly divergent terms  are canceled  {\it sans incertitude}.
       The logarithmically  divergent symmetric  integration (tensor integration) is investigated for obtaining the consistent and gauge invariant result.
      %

\end{abstract}

%
%
\date{\today}
\maketitle

\section{Introduction}
 Relativistic quantum field theory  (QFT) may encounter divergencies in  integration on momentum without finite bound. 
  If the divergence degree is higher than logarithmic, the shift of the integral momentum may lead to change of the 'original result'.
More concretely, for a specific quantity which can be calculated in momentum space with a complete set of  Feynman diagrams to a certain order of the coupling constant, the {\it independent} loop momenta which are integrated to infinity is correlated between these diagrams, and are not free to be set for each single one without referring to the others. If one momentum in   some of the diagram(s) is not correctly assigned, this can correspond to  a shift on this momentum from its correct one for this (these)  diagram(s). The final  result hence can be changed when the corresponding divergence degree is  high.
 One generally can employ  regularization to deal with the divergent terms for the purpose of the proper finite result.
For example, if one starts  from a divergency-free  Lagrangian, as in dimensional  regularization scheme (D.Reg.), the setting of the independent integral momenta could be free from any  ambiguity. The reason is that D.Reg. can be understood as to define a QFT in D-dimension space-time, and in this space-time the theory does not have any divergent integral. 
However, there may be ambiguity in various regularization schemes themselves. And in principle, regularization is not necessary but only a technique for feasibility in practice. Only in the case it will not lead to any ambiguity, it is helpful and useful.
Furthermore,  some regularization schemes are applied to  the 'original' divergent integral of each diagram, but  what is the 'original definition of the divergent integral' for each diagram, even  formally?

We have pointed out that, in four-dimension Minkowski space-time, in which all present  basic physical  QFT's are defined, there exists the 'original definition of the integral' \cite{Li:2017hnv}, {\it at least when only considering the S-matrix theory}.
 This was provided  in Dyson's  original paper \cite{Dyson:1949ha} for S matrix calculation  in momentum space, hence in the following  referred  as 'Dyson scheme'. It also had been   introduced in many QFT text books (see, e.g., \cite{lurie}),  but later was ignored, especially after the D.Reg.  emerged   and dictated most of the practical calculations.
This original definition is not dependent on the regularization and is of no  ambiguity in any of the Feynman diagrams. Starting from this definition, one can explore ways to do the  calculations, e.g.,  those divergences of  high degree   to be cancelled, and  the proper finite or to the most logarithmically   divergent results are obtained. Then, the setting of the independent integral momenta is free from  change by shift.
The definition based on the  Dyson paper is very simple, just as the following when writing down the propagators and vertices for a Feynman  diagram in momentum space:
 Any propagator with momentum $q$ has an extra $[\int \frac{d^4 q}{(2\pi)^4}]$ 'operator',  i.e., should this integration on $q$ to be done in the calculation of the Feynman diagram;
  any vertex has an extra factor $(2\pi)^4 \delta (\sum_i q_i)$ (with  $q_i$,  each momentum of all the propagators attaching to the vertex,    incoming). The latter is just the momentum-energy conservation of each vertex.
  This  means  not to integrate the Dirac $\delta$ function of each vertex  at the beginning.
  Such a  simple essence implies  that  calculation for S-matrix in QFT DOES NOT have the  ambiguity of integral momentum assigning from the starting point. The delta functions are to be integrated only when irrelevant from  introducing    shift of any momentum in the integrand (or shift  allowed for logarithmic divergency or finite).

  This essence can be better understood by recalling the  Feynman propagator in coordinate space  expressed by the momentum integration, which is well defined and only singular on light cone  (see, e.g.,  \cite{bogoliubov}).  In the definition, the   exponential factor $e^{i k^\mu (x_\mu-y_\mu)}$ is very important. It is by integrating the space-time variable in the exponential which is inherent in the calculation of the S-matrix, that one gets the momentum-energy conservation $\delta$ function corresponding to each vertex \cite{Dyson:1949ha} (reflecting the relation of the  momentum-energy conservation and space-time displacement invariance). At the same time, the integration on the momentum  is retained, and we get the 'Feynman Rules'  in momentum space as partially mentioned in the above paragraph. Naturally, the original integral variables, i.e., the momenta which are originated from the expression of each of the propagators in momentum space, are the 'original' integral momenta for each certain single diagram.  Only that these momenta are now  related and restricted to each other by the corresponding $\delta$ functions from the vertices, and this relation is the most important relation in the following calculation, as we will show.
  This also,  in the amplitude,  determines  the functional form of each of the factors coming from the  propagators.

In this paper we apply this Dyson scheme to investigate the axial vector current anomaly (ACA).  %
As is well known,  in ACA calculation each of the one-loop  triangle diagram is linearly divergent, the free shift of integral momentum of each single diagram is not guaranteed, 
  and  one encounters the problem how to assign the independent integral loop momentum of each Feynman diagram.
Alternatively,  when employing  the  D.Reg., one has to  face and solve the '$\gamma_5$'  problem.
 But this is a good arena for Dyson scheme.  Since  the calculation is straightforwardly in four-dimension Minkowski space-time, it is free from  the $\gamma_5$ definition ambiguity. At the same time, since there is no ambiguity to assign the integral momenta in Dyson scheme,   regularization is not needed in the calculation
procedure. The divergencies are naturally cancelled and the definite  finite result is obtained. 
   As the final result is finite, renormalization is not needed.  no extra (physical) renormalization condition is introduced.  In this case the unambiguity of the Dyson scheme is  quite nontrivial.  

         %


In Sec. 2, we write down the vector and axial vector currents to be discussed, and show the Dyson scheme naturally leads to vector current conservation. 
In Sec. 3,  we first demonstrate formally  the  axial vector current conservation  in the Dyson scheme. Then  a detailed calculation  is done, employing only the relations of the loop momenta provided by the $\delta$ function of each vertex.
This calculation finally leads to the logarithmic symmetric (tensoral) integration to be determined.
In Sec. 4, we devote to investigate this logarithmic symmetric integration, especially the surface term and boundary condition, to show how the  self-consistent and gauge invariant result is obtained.
Conclusions and discussions  
 are  in Sec.5.

\section{Definition of  currents and conservation of  vector one}
For a  Dirac field which is the solution of the Dirac equation with  a gauge interaction (We restrict our discussion for this most conventional case only for simplicity),
the  vector current and the axial vector current can be constructed as
$j_\mu=\bar \psi \gamma_\mu \psi$,  $j^A_\mu=\bar \psi \gamma_\mu \gamma_5 \psi$,  respectively (for some purpose we may need an antisymmetrical one).
The current as a whole should be the singlet of the gauge group, and hence summation of the inner quantum number is indicated. 
 From the  Dirac equation, we have
$\partial ^\mu j_\mu=0$  and   $\partial ^\mu j^A_\mu\propto m$. When $m=0$, the axial vector current  also is a conserved current. 
These are formal result from (Dirac) equation of motion, and the derivation is valid for both classical and quantum cases (i.e., Dirac equation of field operators). For the quantum case at tree level, this conservation is respected.
However, It has been pointed out that the conservation of axial  vector current at $m=0$ is broken at loop   level \cite{Adler,Jackiw,Bardeen,Fujikawa}.

Until now, all the calculations to give this conclusion employ some kind of regularization (even the path integral way) because of the    divergencies of mediate steps (e.g., those of the  individual Feynman diagram) \cite{Grange:2020yde}.  
The calculation of the one loop triangle diagrams is the explicit demonstration and is the key point of the study of the  anomaly \cite{AdlerBardeen}.
Each triangle diagram is linearly divergent, the calculation encounters the ambiguity of how to set the independent integral momentum and/or the  ambiguity of the  regularization scheme. 
 %
As we have mentioned, and  will show by the following calculations that the original Feynman  rules for S-matrix of the Dyson scheme are of no ambiguity on the assignment of the integral momenta;  
  and that the cancellation of divergencies between diagrams are as natural result rather than  as   purpose or restriction to direct the calculations.
 %


\begin{figure}[htb]
   	\centering
   \begin{tabular}{cccccc}
   \includegraphics[width=6cm]{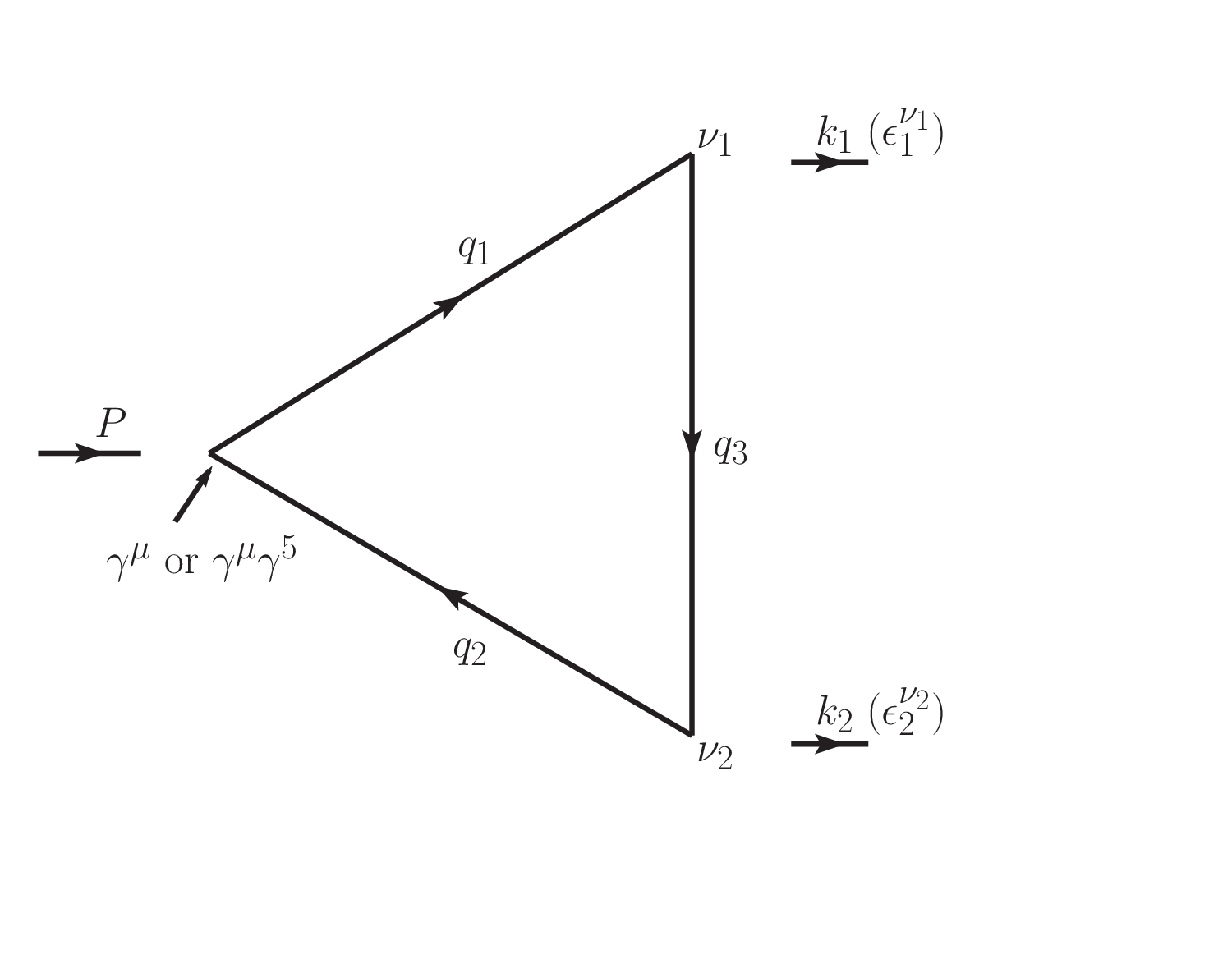}~~~&~~~
   \includegraphics[width=6cm]{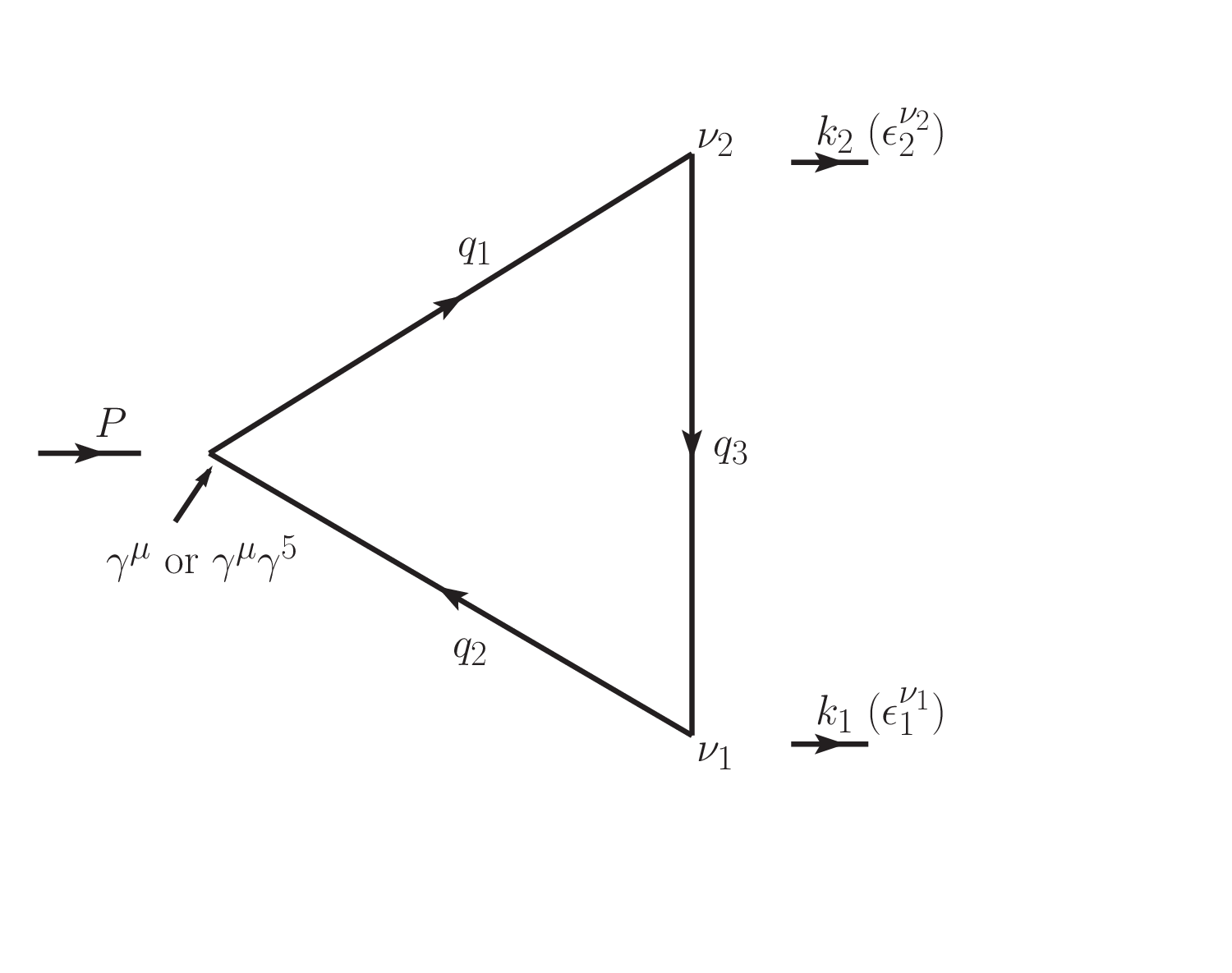}
   \end{tabular}
   	\caption{Two triangle diagrams. The kind and index of each vertex are labeled. These two  diagrams are related   by exchanging  the indices $\nu_1$ with $\nu_2$ and  $k_1$ with $k_2$ simultaneously.}\label{fd}
   \end{figure}

    To concentrate on the key point, we only consider the massless Fermion case, and  start from the amplitudes (see  Fig. 1, also see, e.g., \cite{Ma:2005md}):
\begin{eqnarray}
\label{t1}
 P _\mu \int d^4q_1 d^4q_2 d^4q_3  \delta(q_1-q_2-P) \delta(q_1-q_3-k_1)\delta(q_3-q_2-k_2)
\frac{tr(\gamma^{\nu_1} \slash \hspace{-0.19cm} q_1  \gamma^\mu \gamma^5 \slash \hspace{-0.19cm} q_2 \gamma^{\nu_2} \slash \hspace{-0.19cm} q_3)} {q_1 ^2 q_2 ^2q_3 ^2},
\end{eqnarray}
\begin{eqnarray}
\label{t2}
 P _\mu \int d^4q_1 d^4q_2 d^4q_3  \delta(q_1-q_2-P) \delta(q_1-q_3-k_2)\delta(q_3-q_2-k_1)
\frac{tr(\gamma^{\nu_2} \slash \hspace{-0.19cm} q_1  \gamma^\mu \gamma^5 \slash \hspace{-0.19cm} q_2 \gamma^{\nu_1} \slash \hspace{-0.19cm} q_3)} {q_1 ^2 q_2 ^2q_3 ^2}.
\end{eqnarray}
Each of the above equations corresponds to one of the  two diagrams.    The coupling constant, the overall -1 factor from fermion loop and the i factors are  neglected in the above amplitudes and in following calculations. Here we emphasize that the integral momenta $q_1,  q_2,  q_3$ in these two expressions do not mean the same, since they are just dummy variables. 
As a matter of fact, the cancellation of divergencies happens {\it at integral level not the integrand level}, which is  typical of this Dyson  scheme  and will be seen in the following.
 As convention, the S-matrix and T-matrix have the relation $S=I+i T$, and the matrix element between initial and final states
 $i T_{fi}=i (2\pi)^4 \delta(P_f-P_i) \mathfrak{M}_{fi}$,
for the case without the presence of  any external classical field which  breaks the space-time displacement invariance.
Here  we keep all the  integral momenta,   each  of which corresponds to  one of all the  propagators respectively and hence keep  all the
$\delta$ functions, each  of which respectively corresponds to one of  all the    vertices.  The $\delta$ function corresponding to  the initial-final state energy momentum conservation
is contained in the $\delta$ functions of all vertices. After integrating over them, one will get the above form of T-matrix element with the $\mathfrak{M}_{fi}$
only containing the integration of the independent loop momenta,  without the $\delta$ functions attached to the vertices.
This is the standard procedure in developing the Dyson-Wick perturbation theory in the interaction picture \cite{Dyson:1949ha}.  The four-momentum conservation
$\delta$ function attached to each of the vertices is the result of integration  of space-time contained in the perturbative expansion of the
S-matrix, and is the manifestation of space-time displacement invariance.
Our way to employ the Dyson scheme is that we do not   'integrate out' any of  the $\delta$ functions corresponding to  each of the  vertices until the situation that one of the momenta  does not appear in the integrand except the  $\delta$ functions. 
So here we deal with the matrix elements $T_{fi} $ rather than $\mathfrak{M}_{fi}$, and the
 $(2 \pi )^4$  factors are just cancelled between the integral variable denominator and the factor before each $\delta$ function. 

 It is obvious that,
$\int dx f(x) \delta(x-a)=\int dx f(a) \delta(x-a)$. This means that the relations of different momenta are solid to everywhere in momentum space, until  the boundary at infinity. Any replacement by the  relations $P=q_1-q_2$ and $q_3=q_1-k_1=q_2+k_2$ in the following can be done without   setting any special integral  momenta or doing shifting. So we get, corresponding to the above Eqs. (1) and (2),
\begin{eqnarray}
\label{t1p}
\int d^4q_1 d^4q_2 d^4q_3  \delta(q_1-q_2-P) \delta(q_1-q_3-k_1)\delta(q_3-q_2-k_2)
\frac{tr(\gamma^{\nu_1} \slash \hspace{-0.19cm} q_1 (\slash \hspace{-0.19cm}q_1-\slash \hspace{-0.19cm}q_2)  \gamma^5 \slash \hspace{-0.19cm} q_2 \gamma^{\nu_2} \slash \hspace{-0.19cm} q_3)} {q_1 ^2 q_2 ^2q_3 ^2},
\end{eqnarray}
\begin{eqnarray}
\label{t2p}
 \int d^4q_1 d^4q_2 d^4q_3  \delta(q_1-q_2-P) \delta(q_1-q_3-k_2)\delta(q_3-q_2-k_1)
\frac{tr(\gamma^{\nu_2} \slash \hspace{-0.19cm} q_1  (\slash \hspace{-0.19cm}q_1-\slash \hspace{-0.19cm}q_2) \gamma^5 \slash \hspace{-0.19cm} q_2 \gamma^{\nu_1} \slash \hspace{-0.19cm} q_3)} {q_1 ^2 q_2 ^2q_3 ^2}.
\end{eqnarray}
 It looks as if some terms can be quadratic divergent.  But  we will show  them to the most linear.

However, before exploring  the axial vector, we first get the result  that the vector current is naturally conserved.


The vector current conservation is to calculate the expressions which  can be obtained from Eqs., (1,2) or (3,4) with the vertex $\gamma^\mu \gamma^5$ replaced by  $\gamma^\mu $:
 \begin{eqnarray}
\label{v1}
\int d^4q_1 d^4q_2 d^4q_3  \delta(q_1-q_2-P) \delta(q_1-q_3-k_1)\delta(q_3-q_2-k_2)
\frac{tr(\gamma^{\nu_1} \slash \hspace{-0.19cm} q_2 \gamma^{\nu_2} \slash \hspace{-0.19cm} q_3)} {q_2 ^2q_3 ^2}\\
\label{v2}
-\int d^4q_1 d^4q_2 d^4q_3  \delta(q_1-q_2-P) \delta(q_1-q_3-k_1)\delta(q_3-q_2-k_2)\frac{tr(\gamma^{\nu_1} \slash \hspace{-0.19cm} q_1 \gamma^{\nu_2} \slash \hspace{-0.19cm} q_3)} {q_1 ^2q_3 ^2},
\end{eqnarray}
\begin{eqnarray}
\label{v3}
\int d^4q_1 d^4q_2 d^4q_3  \delta(q_1-q_2-P) \delta(q_1-q_3-k_2)\delta(q_3-q_2-k_1)
\frac{tr(\gamma^{\nu_2} \slash \hspace{-0.19cm} q_2 \gamma^{\nu_1} \slash \hspace{-0.19cm} q_3)} {q_2 ^2q_3 ^2}\\
\label{v4}
-\int d^4q_1 d^4q_2 d^4q_3  \delta(q_1-q_2-P) \delta(q_1-q_3-k_2)\delta(q_3-q_2-k_1)\frac{tr(\gamma^{\nu_2} \slash \hspace{-0.19cm} q_1 \gamma^{\nu_1} \slash \hspace{-0.19cm} q_3)} {q_1 ^2q_3 ^2}.
\end{eqnarray}
Now in each term of Eqs. (5,7) ((6,8)),  $q_1$  ($q_2$) only appears in the delta functions, so can be integrated to eliminate, and we  get
\begin{eqnarray}
\label{v1p}
\int d^4q_2 d^4q_3  \delta(q_3-q_2+k_1-P) \delta(q_3-q_2-k_2)
\frac{tr(\gamma^{\nu_1} \slash \hspace{-0.19cm} q_2 \gamma^{\nu_2} \slash \hspace{-0.19cm} q_3)} {q_2 ^2q_3 ^2}\\
-\int d^4q_1  d^4q_3  \delta(q_1-q_3+k_2-P) \delta(q_1-q_3-k_1)\frac{tr(\gamma^{\nu_1} \slash \hspace{-0.19cm} q_1 \gamma^{\nu_2} \slash \hspace{-0.19cm} q_3)} {q_1 ^2q_3 ^2},
\end{eqnarray}
\begin{eqnarray}
\label{v2p}
\int  d^4q_2 d^4q_3  \delta(q_3-q_2+k_2-P) \delta(q_3-q_2-k_1)
\frac{tr(\gamma^{\nu_2} \slash \hspace{-0.19cm} q_2 \gamma^{\nu_1} \slash \hspace{-0.19cm} q_3)} {q_2 ^2q_3 ^2}\\
-\int d^4q_1  d^4q_3  \delta(q_1-q_3+k_1-P) \delta(q_1-q_3-k_2)\frac{tr(\gamma^{\nu_2} \slash \hspace{-0.19cm} q_1 \gamma^{\nu_1} \slash \hspace{-0.19cm} q_3)} {q_1 ^2q_3 ^2}.
\end{eqnarray}
Here we see the terms Eq.(9) with Eq.(12)  as well as  Eq.(10) with  Eq.(11),  are just negative to each other {\it at integral level}.  So the summation must be zero.
To see this more clearly, one can replace the dummy variables $q_3, q_2$  in (9) and  (11) by $q'_1, q'_3$ respectively.
%
We see the very advantage of this Dyson scheme is that the cancellation is on the {\it integral level but not simply} the integrand level.

We also would like to address, employing the relations provided by the $\delta$ functions, we can simplify/derive the above zero result in many ways, e.g.,  as we will done on the following axial vector current case.  The above one is the most simple, which relies on the symmetric
property of the trace of the one loop triangle diagram on $\nu_1$ and $\nu_2$.  

At the same time, one can observe that this is a simple application of the  Furry theorem.  According to the Furry theorem, the summation of these two diagrams, even without contracting with the momentum $P^\mu$, should be zero. But explicit calculation implies that, this simplest case of Furry theorem, i.e., this triangle diagrams with three vertices, can be non-zero if the integral momenta are not properly set.  The reason is of course the existence of the linear divergency of each diagram.
  However, without assigning the independent integral momentum, but straightforwardly starting from the original Feynman rules of the Dyson scheme, one can demonstrate that the Furry theorem is true for this case. Then by applying the Furry theorem, we conclude that to the   triangle diagram level,  vector current must be conserved.
      To employ the above Feynman rules without integrating the $\delta$ functions, and to employ the charge conjugation operator to demonstrate that the summation of these two diagrams equals to zero, can be found in many  QFT textbooks.

 \section{Axial vector current  in  Dyson scheme}
 Now let's turn back to the axial vector current case.
A simple demonstration that the summation of two triangle diagrams  is zero, i.e., anomaly-free, can be done by the similar way as the  above derivation for vector current:
From Eqs. (3), (4)
\begin{eqnarray}
\label{a1}
-\int d^4q_1 d^4q_2 d^4q_3  \delta(q_1-q_2-P) \delta(q_1-q_3-k_1)\delta(q_3-q_2-k_2)
\frac{tr(\gamma^5 \gamma^{\nu_1} \slash \hspace{-0.19cm} q_2 \gamma^{\nu_2} \slash \hspace{-0.19cm} q_3)} {q_2 ^2q_3 ^2}\\
+\int d^4q_1 d^4q_2 d^4q_3  \delta(q_1-q_2-P) \delta(q_1-q_3-k_1)\delta(q_3-q_2-k_2)\frac{tr(\gamma^5 \gamma^{\nu_1} \slash \hspace{-0.19cm} q_1 \gamma^{\nu_2} \slash \hspace{-0.19cm} q_3)} {q_1 ^2q_3 ^2}\nonumber,
\end{eqnarray}
\begin{eqnarray}
\label{a2}
-\int d^4q_1 d^4q_2 d^4q_3  \delta(q_1-q_2-P) \delta(q_1-q_3-k_2)\delta(q_3-q_2-k_1)
\frac{tr(\gamma^5 \gamma^{\nu_2} \slash \hspace{-0.19cm} q_2 \gamma^{\nu_1} \slash \hspace{-0.19cm} q_3)} {q_2 ^2q_3 ^2}\\
+\int d^4q_1 d^4q_2 d^4q_3  \delta(q_1-q_2-P) \delta(q_1-q_3-k_2)\delta(q_3-q_2-k_1)\frac{tr(\gamma^5 \gamma^{\nu_2} \slash \hspace{-0.19cm} q_1 \gamma^{\nu_1} \slash \hspace{-0.19cm} q_3)} {q_1 ^2q_3 ^2}\nonumber.
\end{eqnarray}
Now as the vector current case, in each term of the above two expressions,  $q_1$ or $q_2$ only appears in the $\delta$ functions, so can be  eliminated by integration. We  get
\begin{eqnarray}
-\int d^4q_2 d^4q_3  \delta(q_3-q_2+k_1-P) \delta(q_3-q_2-k_2)
\frac{tr(\gamma^5 \gamma^{\nu_1} \slash \hspace{-0.19cm} q_2 \gamma^{\nu_2} \slash \hspace{-0.19cm} q_3)} {q_2 ^2q_3 ^2} \label{a1p}\\
+\int d^4q_1  d^4q_3  \delta(q_1-q_3+k_2-P) \delta(q_1-q_3-k_1)\frac{tr(\gamma^5\gamma^{\nu_1} \slash \hspace{-0.19cm} q_1 \gamma^{\nu_2} \slash \hspace{-0.19cm} q_3)} {q_1 ^2q_3 ^2}\label{a1p2},
\end{eqnarray}
\begin{eqnarray}
-\int  d^4q_2 d^4q_3  \delta(q_3-q_2+k_2-P) \delta(q_3-q_2-k_1)
\frac{tr(\gamma^5 \gamma^{\nu_2} \slash \hspace{-0.19cm} q_2 \gamma^{\nu_1} \slash \hspace{-0.19cm} q_3)} {q_2 ^2q_3 ^2}\label{a2p} \\
+\int d^4q_1  d^4q_3  \delta(q_1-q_3+k_1-P) \delta(q_1-q_3-k_2)\frac{tr(\gamma^5 \gamma^{\nu_2} \slash \hspace{-0.19cm} q_1 \gamma^{\nu_1} \slash \hspace{-0.19cm} q_3)} {q_1 ^2q_3 ^2}\label{a2p2}.
\end{eqnarray}
 Here we see that the terms (\ref{a1p}) with (\ref{a2p2}),   as well as  (\ref{a1p2}) with  (\ref{a2p}),  are just negative to each other {\it at integral level}.  So the summation must be zero \footnote{Employing the way to prove the Furry theorem, i.e., employing the charge conjugation operator, one finds these two diagrams equal to each other. Combined with the above zero result of the summation, this may indicate that the $\gamma_5$ in fact inherently
nullize each diagram, which only {\it superficially } appears as infinite.}.
Again, we see the advantage of this scheme, the cancellation is on the {\it integral level but not simply} the integrand level.

On the other hand,  we 
take a more 'complex' way to do the derivation, to see some subtle elements for this anomaly-free result.
From Eqs. (1) and  (3),
\begin{eqnarray}
\label{act1}
T_1 & = & 4 \int d^4q_1 d^4q_2 d^4q_3  \delta(q_1-q_2-P) \delta(q_1-q_3-k_1)\delta(q_3-q_2-k_2) \\ \nonumber
 ~ & \times &\frac{-q_1^2\epsilon^{\nu_{1}\alpha \nu_{2}\beta}q_{2\alpha}q_{3\beta}+q_2^2\epsilon^{\nu_{1}\alpha \nu_{2}\beta}q_{1\alpha}q_{3\beta}} {q_1 ^2 q_2 ^2q_3 ^2}.
\end{eqnarray}
 From eqs. (2) and  (4),
 \begin{eqnarray}
\label{act2}
T_2 & = & 4 \int d^4q_1 d^4q_2 d^4q_3  \delta(q_1-q_2-P) \delta(q_1-q_3-k_2)\delta(q_3-q_2-k_1) \\ \nonumber
 ~ & \times & \frac{-q_1^2\epsilon^{\nu_{2}\alpha \nu_{1}\beta}q_{2\alpha}q_{3\beta}+q_2^2\epsilon^{\nu_{2}\alpha \nu_{1}\beta}q_{1\alpha}q_{3\beta}} {q_1 ^2 q_2 ^2q_3 ^2}.
\end{eqnarray}
 The above $T_1$ and $T_2$ are tensors and have two Minkowski indices $\nu_1, \nu_2$ to be contracted with two photon polarization vectors, respectively.
At first look, it seems  quadratically divergent, but   to the most     linearly divergent. The reason is simply that to contract with the totally asymmetric tensor, those four momenta  must be different to each other for non-zero result.

 The way to cancel the linear divergency  is between these two diagrams.
 We employ the relations   $q_1=q_3+k_1$, $q_2=q_3-k_2$, and relations like $\epsilon^{\nu_{1}\alpha \nu_{2}\beta}q_{2\alpha}q_{2\beta}=0$:
 \begin{eqnarray}
\label{act1p}
T_1 & = & 4 \int d^4q_1 d^4q_2 d^4q_3  \delta(q_1-q_2-P) \delta(q_1-q_3-k_1)\delta(q_3-q_2-k_2)  \nonumber \\
 ~ & \times &\frac{-(q_3^2+2 q_3 \cdot k_1)\epsilon^{\nu_{1}\alpha \nu_{2}\beta}q_{2\alpha}k_{2\beta}+(q_3^2-2 q_3 \cdot k_2)(-\epsilon^{\nu_{1}\alpha \nu_{2}\beta}q_{1\alpha}k_{1\beta})} {q_1 ^2 q_2 ^2q_3 ^2}.
\end{eqnarray}
By this way we can see that $T_1$ is to the most linearly divergent, and can be separated in to two parts, one is linear and the other is logarithmic, $T_1=TLI_1+TLG_1$.  This way also helps us to get a result without $q_3$ in the linear-divergent integrand.
\begin{eqnarray}
\label{li1}
TLI_1 & = & 4 \int d^4q_1 d^4q_2 d^4q_3  \delta(q_1-q_2-P) \delta(q_1-q_3-k_1)\delta(q_3-q_2-k_2)  \nonumber \\
 ~ & \times &\frac{-\epsilon^{\nu_{1}\alpha \nu_{2}\beta}q_{2\alpha}k_{2\beta}-\epsilon^{\nu_{1}\alpha \nu_{2}\beta}q_{1\alpha}k_{1\beta}} {q_1 ^2 q_2 ^2} \nonumber \\
 ~ & = & 4 \int d^4q_1 d^4q_2   \delta(q_1-q_2-P) \delta(q_1-q_2-k_1-k_2)  \nonumber \\
 ~ & \times &\frac{-\epsilon^{\nu_{1}\alpha \nu_{2}\beta}q_{2\alpha}k_{2\beta}-\epsilon^{\nu_{1}\alpha \nu_{2}\beta}q_{1\alpha}k_{1\beta}} {q_1 ^2 q_2 ^2}.
\end{eqnarray}
The second equation is obtained just because $q_3$  cancelled between numerator and denominator in the integrand, and only appears in $\delta$ functions. So  $q_3$ can be integrated and eliminated.  $TLI_1$ can be combined with corresponding terms in $TLI_2$ in the following. On the other  hand,
 \begin{eqnarray}
\label{log1}
TLG_1 & = & 4 \int d^4q_1 d^4q_2 d^4q_3  \delta(q_1-q_2-P) \delta(q_1-q_3-k_1)\delta(q_3-q_2-k_2)  \nonumber \\
 ~ & \times &\frac{-2 q_3 \cdot k_1\epsilon^{\nu_{1}\alpha \nu_{2}\beta}q_{2\alpha}k_{2\beta}+2 q_3 \cdot k_2\epsilon^{\nu_{1}\alpha \nu_{2}\beta}q_{1\alpha}k_{1\beta}} {q_1 ^2 q_2 ^2q_3 ^2}.
\end{eqnarray}

 Correspondingly, similar can be done on $T_2=TLI_2+TLG_2$, using relations from the $\delta$ functions, $q_1=q_3+k_2$, $q_2=q_3-k_1$, as well as the property of  the totally antisymmetric tensor:
 \begin{eqnarray}
\label{act2p}
T_2 & = & 4 \int d^4q_1 d^4q_2 d^4q_3  \delta(q_1-q_2-P) \delta(q_1-q_3-k_2)\delta(q_3-q_2-k_1) \nonumber \\
 ~ & \times & \frac{-(q_3^2+2 q_3 \cdot k_2)\epsilon^{\nu_{2}\alpha \nu_{1}\beta}q_{2\alpha}k_{1\beta}+(q_3^2-2 q_3 \cdot k_1)(-\epsilon^{\nu_{2}\alpha \nu_{1}\beta}q_{1\alpha}k_{2\beta})} {q_1 ^2 q_2 ^2q_3 ^2}.
\end{eqnarray}
Hence
\begin{eqnarray}
\label{act2p}
TLI_2 & = & 4 \int d^4q_1 d^4q_2 d^4q_3  \delta(q_1-q_2-P) \delta(q_1-q_3-k_2)\delta(q_3-q_2-k_1)  \nonumber \\
 ~ & \times & \frac{-\epsilon^{\nu_{2}\alpha \nu_{1}\beta}q_{2\alpha}k_{1\beta}-\epsilon^{\nu_{2}\alpha \nu_{1}\beta}q_{1\alpha}k_{2\beta}} {q_1 ^2 q_2 ^2} \nonumber \\
 & = & 4 \int d^4q_1 d^4q_2  \delta(q_1-q_2-P) \delta(q_1-q_2-k_1-k_2)  \nonumber \\
 ~ & \times & \frac{-\epsilon^{\nu_{2}\alpha \nu_{1}\beta}q_{2\alpha}k_{1\beta}-\epsilon^{\nu_{2}\alpha \nu_{1}\beta}q_{1\alpha}k_{2\beta}} {q_1 ^2 q_2 ^2}  \nonumber \\
 & = & 4 \int d^4q_1 d^4q_2  \delta(q_1-q_2-P) \delta(q_1-q_2-k_1-k_2)  \nonumber \\
 ~ & \times & \frac{\epsilon^{\nu_{1}\alpha \nu_{2}\beta}q_{2\alpha}k_{1\beta}+\epsilon^{\nu_{1}\alpha \nu_{2}\beta}q_{1\alpha}k_{2\beta}} {q_1 ^2 q_2 ^2}.
\end{eqnarray}
\begin{eqnarray}
\label{act2p}
TLG_2 & = & 4 \int d^4q_1 d^4q_2 d^4q_3  \delta(q_1-q_2-P) \delta(q_1-q_3-k_2)\delta(q_3-q_2-k_1)  \nonumber \\
 ~ & \times & \frac{-2 q_3 \cdot k_2\epsilon^{\nu_{2}\alpha \nu_{1}\beta}q_{2\alpha}k_{1\beta}+2 q_3 \cdot k_1\epsilon^{\nu_{2}\alpha \nu_{1}\beta}q_{1\alpha}k_{2\beta}} {q_1 ^2 q_2 ^2q_3 ^2}.
\end{eqnarray}

Now it is clear that in the  summation of the {\it integral} $TLI_1$ and $TLI_2$,  the integrand can  be summed to get  (corresponding terms in these two integrands: $1_1 \& 2_2, 2_1 \&1_2)$, with $q_1-q_2=k_1+k_2$,
\begin{equation}
LG_3=TLI_1+TLI_2= 4 \int d^4q_1 d^4q_2   \delta(q_1-q_2-P) \delta(q_1-q_2-k_1-k_2) \frac{2 \epsilon^{\nu_{1}\alpha \nu_{2}\beta}k_{1\alpha}k_{2\beta}} {q_1 ^2 q_2 ^2}.
 \end{equation}

 Now the summation of the two diagrams of the axial vector current is $TLG_1+TLG_2+LG_3$. It is logarithmic, and can be treated  with  the standard way without the need of any regularization.   The calculations can be done all in four-dimension Minkowski space-time. One  can  use Feynman-Schwinger parameterization, shift, and  Wick rotation to  show that the logarithmic terms are cancelled by symmetric integration and get the final result (see next section).    

In the above derivation, $k_1^2$ and $k_2^2$ do not appear in linear divergent terms (in fact do  not appear in all divergent terms). So our discussion on the cancellation of high degrees of  divergencies is of no  relevance with the physical
condition of the momenta $k_1^2$ and $k_2^2$. However, we would like to address again that the momentum conservation of each vertex is intrinsic in the Feynman rules, which plays the central r\^{o}le in the above derivation.


\section{Logarithmic symmetric (tensor) integration and  axial vector current anomaly}
The calculation on  $TLG_1+TLG_2+LG_3$, 
 algebrically a little longer than the above,  is quite standard and conventional.  So we will not put the details here.  We write down the details in above sections  just  because they are not normal. Here for the  calculation on  $TLG_1+TLG_2+LG_3$, only the following key point has to be discussed,
which is pointed out long ago but not full analyzed:  After the Feynman-Schwinger parameterization, and the shift is made (which  will not change the result \cite{jr}),  the logarithmically divergent symmetric integration
\begin{equation}
\int d^4 l \frac{l^\mu l^\nu}{(l^2-\Delta)^3},
\end{equation}
are to be calculated by the following way, and the boundary condition for issues such as Feynman propagators in Dyson scheme has to be explicitly employed.
We start from
\begin{equation}
\partial_l^\mu \frac{l^\nu}{(l^2-\Delta)^2}=\frac{g^{\mu \nu}}{(l^2-\Delta)^2}-\frac{4 l^\mu l^\nu}{(l^2-\Delta)^3}=\frac{l^2 g^{\mu \nu}- 4 l^\mu l^\nu}{(l^2-\Delta)^3}-\frac{\Delta g^{\mu \nu}}{(l^2-\Delta)^3}.
\end{equation}
So, we have
\begin{equation}
\int d^4 l \frac{l^2 g^{\mu \nu}- 4 l^\mu l^\nu}{(l^2-\Delta)^3}= \int d^4 l \partial_l^\mu \frac{l^\nu}{(l^2-\Delta)^2}+ \int d^4 l \frac{\Delta g^{\mu \nu}}{(l^2-\Delta)^3}.
\end{equation}
In the  above equation, employing the Gaussian theorem in 4-dimension Minkowski (phase) space, the first term at the  right-hand side is
\begin{equation}
\label{surf}
\int d^4 l \partial_l^\mu \frac{l^\nu}{(l^2-\Delta)^2} = \int d\sigma^\mu \frac{l^\nu}{(l^2-\Delta)^2}.
\end{equation}
$d\sigma^\mu$ is the differential element of the  three-dimension hypersurface  as the boundary of the whole four-dimension Minkowski phase space, the conjugation of the four-dimension Minkowski spacetime  in which the Dyson perturbative theory is defined.
As the Gaussian theorem allows, the three-dimension hypersurface boundary can be as far away as possible.
Then this surface integral Eq. (\ref{surf}) equals to zero, which is a nontrivial conclusion by the definition  of the free Feynman propagator adopted by the Dyson perturbative theory.

As we all know, the Feynman  propagator is defined as a contour integral of the complex $k^0$ plane. To have a definite result taking into account the poles, one must require $k^0 >|\vec{k}|$ (In fact we generally take the contour with $k^0 /|\vec{k}| \to \infty$). However, for employing the Gaussian theorem and to be consistent
with  this boundary condition of the  free Feynman propagator, one can require a boundary in infinity and with the $k^0$ much much larger than $|\vec{k}|$.  The reason is that the  hypersurface of the Gaussian theorem can be much much  larger than 'necessary', with only requirement that the volume in the hypersurface covers the whole integration volume.  So on such a hypersurface the integrand is vanishing and we get the zero result of the Eq. (\ref{surf}). This also precisely define the boundary condition of the Dyson scheme.  We can put the boundary hypersurface like this because the 'elements' in  perturbative calculation in the Dyson scheme of the interaction picture \footnote{More precisely, the 'renormalized interaction picture', see, P. T. Matthews and A. Salam,  Phys. Rev. 94 (1954)185.} is the free theory (propagator/Green function), free spacetime, free vacuum, and the scattering theory is a static solution. The asymptotic  solution for a 'single' particle is described as the summation of a plane wave and a sphere wave, as the  solution of the homogeneous and inhomogeneous Helmholtz Equations, respectively.  In this standard case we can put the $|\vec{k}|$ as large as possible, also put $k^0$ as large as possible, while requiring $k^0$ much much larger than $|\vec{k}|$ (This seems consistent with the contour with $k^0 /|\vec{k}| \to \infty$, which not changed by Wick rotation). Hence the 'most far'---especially most far for $k^0$--- hypersurface integral as right-hand side of  Eq. (\ref{surf})  can be zero.

By the above discussion, we see in four-dimension Minkowski phase space,
\begin{equation}
\int d^4 l \frac{l^2 g^{\mu \nu}- 4 l^\mu l^\nu}{(l^2-\Delta)^3}=  \int d^4 l \frac{\Delta g^{\mu \nu}}{(l^2-\Delta)^3}=- i \frac{\pi ^2}{2} g^{\mu \nu},
\end{equation}
or, equally the tensor integration written as
\begin{equation}
\label{tensor}
\int d^4 l \frac{l^\mu l^\nu}{(l^2-\Delta)^3}= \frac{1}{4} \int d^4 l \frac{ g^{\mu \nu}}{(l^2-\Delta)^2}.
\end{equation}
  %
     This equation guarantees the zero result of $TLG_1+TLG_2+LG_3$ in Dyson scheme.
This result gives an explicit example of the difference between Minkowski space from the Euclid one. For the latter,  one may take the most simple periodic boundary condition with the time component $k^0$ at the same footing as the space components, and hence a straightforward replacement $l^\mu l^\nu \to \frac{1}{4} l^2 \delta^{\mu \nu}$ can be taken. But this does not respect the relation $k^0 > |\vec{k}|$. This relation, as explained in the above paragraph, must be respected since our standard scattering theory of the Dyson scheme is a four-dimension covariant (Minkowski) realization of the stationary solution of the Helmholtz Equation.

There is another very crucial point  relevant with this symmetric integration. The axial vector current, or the  summation of these two triangle diagrams without contracted with the momentum on the axial vertex  (e.g., without contracted to $P^\mu$ in Eqs. (1, 2)), should respect the  U(1) (vector) gauge invariance. I.e., when the   polarization vector   contracted with the corresponding vector vertex index is replaced by the corresponding photon four momentum, the result should be zero.  
This can be checked to be true in Fig. 1 on the vertices $\nu_1$ or $\nu_2$.  There are again two ways. The simpler is similar as Eqs. (15-18).
But one can also do more complex calculations as above, replacing the  photon four momentum by the loop momenta  (e.g., in Fig.1 to replace $k_1$ by $q_1-q_3$. This can be done in each diagram guided by two $\delta$ functions for the photon coupling vertices.). Again, the symmetric integration (\ref{tensor}) is necessary  for the gauge invariance.

 Hence, in the Dyson scheme,  straightforward in four dimension Minkowski spacetime, without regularization,   via the formal way or via the detailed calculation employing the symmetric (tensor) integral investigated above, one can get the situation that all three currents, one axial vector and two vector ones, consistently conserve at the same time.

 As a matter of fact, the above investigation is restricted   in only the case of   the S-matrix theory of QFT in free vacuum, i.e., stationary solution of scattering theory in four-dimension Minkowski spacetime \cite{Dyson:1949ha}, rather than any other problems. In general,  QFT models are dynamics models and each is characteristic of a group of partial differential equations (equivalently, Lagrangian), which can be applied to diverse kinds of problems besides the stationary S-Matrix theory.  Each  corresponds to a certain specific physical conditions,  e.g., boundary conditions, and could be represented in terms of specific Lorentz invariant  Green functions for the specific problem. The S-matrix problem in free vacuum is only one of many possible cases of solutions of QFT models, and is characteristic of the free Feynman propagators as Green functions. So that the claim 'the Dyson scheme is unambiguous' here only is 
  restricted to the  S-Matrix theory. 
  %
    %
   Hence the coventional  ACA  calculations with various methods of regularization \cite{Adler,Jackiw,Bardeen,Fujikawa}, (maybe) supplementary boundary conditions or physical inputs to eliminate any uncertainties,   
    are not restricted to be  
   understood as within   the Dyson scheme.
 For a large amount of physical problems which can take the Euclidean spacetime  solution, such as the strong CP problem with an Euclidean soliton solution, one can adopt the na\"ive symmetric integration with the replacement $l^\mu l^\nu=l^2 g^{\mu \nu}$, employing the Euclidean boundary condition. Then one can get the ACA result, with  $TLG_1+TLG_2+LG_3=\frac{1}{4 \pi^2}\epsilon^{\nu_1 \nu_2 \rho \delta} k_{1\rho}k_{2\delta}$.

\section{Conclusion and Discussion}
 In conclusion, in this paper we demonstrate that  the most original Dyson scheme is  definite. The Feynman diagrams can be calculated in four-dimension Minkowski phase space without  regularization.     Gauge invariance is respected.
 The ACA can be calculated  without ambiguity.  It can also relates  with a certain definite boundary condition in momentum space.
  For this purpose the  logarithmically divergent symmetric (tensoral) integration is clarified.  
    We clearly connect this integral with the boundary condition on the hypersurface  at infinity in the Minkowski phase space. 
    The definition of the free Feynman propagator in Minkowski phase space can determine the boundary condition and this also means that (effective)space-time properties other than Minkowski can lead to other kind of boundary condition and hence physics.
    This idea  can be applied to various problems (modeling or calculation).
As a simple example in calculation, 
 the well known 'change by shift in a linear integral' of Jauch and Rohrlich book \cite{jr} 
\begin{equation}
\int d^4k \frac{k^\mu \gamma_\mu}{((k-p)^2+a)^2}-\int d^4k \frac{k^\mu \gamma_\mu+p^\mu \gamma_\mu}{(k^2+a)^2},
\end{equation}
equals to
\begin{equation}
\label{s1}
-\int d^4k \frac{p^\mu \gamma_\mu}{(k^2+a)^2}+\int d^4k k^\mu \gamma_\mu (\frac{1}{((k-p)^2+a)^2}- \frac{1}{(k^2+a)^2}).
\end{equation}
By only momentum shift (since to the most logarithmic) we can get it as
\begin{equation}
-\int d^4k \int dz p_\nu \partial^\nu_k \frac{k^\mu \gamma_\mu}{(k^2+a+p^2 z (1-z))^2}.
\end{equation}
To get the dz integral we have employed Eq. (A5-13) of Jauch and Rohrlich   \cite{jr} for Eq. (\ref{s1}).
The above integral of the total differential again can be written as surface integral by Gaussian theorem.  In this way we need  not  employ the symmetric integration  as Jauch and Rohrlich ($k^\mu k^\nu \to \frac{1}{4}k^2 g^{\mu \nu} $,  corresponding to the case the surface integral not vanishing),
hence eliminate any uncertainty. The concrete value can be obtained at specified boundary condition for the concrete surface integral value. Especially when we adopt the surface integral vanishing, we can get that the shift will NOT change this linear integral. This is an example of the relation of surface integral vanishing and the momentum routine invariance, as pointed by \cite{Ferreira:2011cv}.
This kind of boundary condition is  not restricted to the logarithmic though not able to solve all high order divergence cancellation problems. As another example which also can be found from  \cite{Ferreira:2011cv}, the quadratic case, which appears, e.g., in the well known one-loop QED photon self-energy. The latter has the reputation that there are in history many ways to 'solve' the correct form by considering gauge invariance and seems 'having to' employ a regularization which respect gauge invariance. However, one can easily check by rewriting the superficial quadratic terms into the surface form in four dimension Minkowski phase space, one can always get the correct gauge invariant logarithmic form by requiring surface integral vanishing.


 \section*{Acknowledgments}

 S-Y Li is in debt to Prof. T. T.  Wu for initializing related studies which lead to the investigation on the application of the Dyson scheme, as well as for  many encouraging and instructive  discussions and suggestions including  his early investigation on ABJ triangle diagrams with the most symmetric loop momentum.
  This work is supported in part by National Natural Science Foundation of China (grant Nos. 11635009, 11775130, 12275157) and the Natural Science Foundation of Shandong Province (grant No. ZR2020MA094).


\begin{thebibliography}{150}


\bibitem{Li:2017hnv}
S.~Y.~Li, Z.~G.~Si and X.~F.~Zhang,
[arXiv:1705.04941 [hep-ph]].


\bibitem{Dyson:1949ha}
  F.~J.~Dyson,
  Phys.\ Rev.\  {\bf 75} (1949) 1736.
  doi:10.1103/PhysRev.75.1736

  \bibitem{lurie} David Lurie, Particles and Fields  Inter-science (Wiley), New York, 1968.

\bibitem{bogoliubov}  Bogoliubov, N.N. and Shirkov, D.V. (1959) Introduction to the Theory of Quantized Fields. 2nd Printing Edition, Interscience Publishers, 35.


\bibitem{Adler}
S. L. Adler, Phys. Rev. 177, 2426 (1969).

\bibitem{Jackiw}
J. S. Bell and R. Jachiw,
Il Nuovo Cimento, LX A, 47 (1969).

\bibitem{Bardeen}
W. A. Bardeeen, Phys. Rev. 184, 1848 (1969).

\bibitem{Fujikawa}
K. Fujikawa, Phys. Rev. Lett. 42, 1195 (1979).

\bibitem{Grange:2020yde}
Besides the above mentioned classical papers, a recent example is,
P.~Grang\'e, J.~F.~Mathiot and E.~Werner,
Int. J. Mod. Phys. A \textbf{35} (2020) no.05, 2050025
doi:10.1142/S0217751X20500256
[arXiv:2002.01191 [hep-ph]].

\bibitem{AdlerBardeen}
S. L. Adler and W.A. Bardeen, Phys. Rev. 182, 1517 (1969).


\bibitem{Ma:2005md}
Y.~L.~Ma and Y.~L.~Wu,
Int. J. Mod. Phys. A \textbf{21} (2006), 6383-6456
  doi:10.1142/S0217751X0603309X
  [arXiv:hep-ph/0509083 [hep-ph]].


\bibitem{jr}
J. M. Jauch and F. Rohrlich, The Theory of Photons and Electrons, Springer, 1976. ISBN: 978-3-642-80951-4.



\bibitem{Ferreira:2011cv}
L.~C.~Ferreira, A.~L.~Cherchiglia, B.~Hiller, M.~Sampaio and M.~C.~Nemes,
Phys. Rev. D \textbf{86} (2012), 025016
doi:10.1103/PhysRevD.86.025016
[arXiv:1110.6186 [hep-th]].


\end{thebibliography}
\end{document}